# Two-photon ionization of the *K-shell* of a heavy beryllium-like atomic ion

Hopersky A.N., Nadolinsky A.M., Novikov S.A. *and* Koneev R.V.

Rostov State Transport University, 344038, Rostov-on-Don, Russia
E-mail: qedhop@mail.ru, amnrnd@mail.ru, sanovikov@gmail.com, koneev@gmail.com

---

**Abstract.** Within the framework of the second order of the non-relativistic quantum perturbation theory, the analytical structure and absolute values of the generalized cross-sections of the two-photon resonant single ionization of the K-shell of heavy beryllium-like ions of titanium ($Ti^{18+}$), chromium ($Cr^{20+}$), iron ($Fe^{22+}$) and zinc ($Zn^{26+}$) atoms were predicted. The complete wave functions of the ground state of the ion and the states of its ionization are obtained in the single-configuration Hartree-Fock approximation. The effects of (a) the occurrence of giant resonances in the subthreshold region of the generalized ionization cross-sections, (b) destructive quantum interference of the probability amplitudes of radiation transitions to intermediate states of *p*-symmetry, and (c) the leading role of the *d*-symmetry of the final ionization state in determining the values of the total generalized cross-section in the region of energies of absorbed photons of the hard X-ray range were established. A scheme of the proposed experiment with linearly polarized X-ray photons is presented to verify the theoretical results obtained.

---

## 1. Introduction

Two-photon (non-linear) ionization of deep shells of atoms, atomic ions, molecules and solids is one of the fundamental processes of the microcosm. With the creation of the x-ray free-electron laser (XFEL) [1] as a source of hard multiphoton radiation, the possibility of high-precision experimental and theoretical studies of this process has opened up (see reviews [2, 3]). Theoretical studies of this process have revealed, in particular, the important role of relativistic effects, shielding effects, and non-dipole (quadrupole) effects in (a) determining the non-resonant structure of the generalized cross-sections of the two-photon ionization of the K-shell of the atom and the atomic ion [4–7] and (b) the angular distribution of photoelectrons produced by the suprathreshold two-photon ionization of the K-shell of the atom [8,9]. In these works, there are no studies of the resonant subthreshold structure of the generalized cross-section of two-photon ionization. Such a structure within the framework of the second-order non-relativistic quantum perturbation theory was first theoretically studied in the works of the authors [10,11] on the example of light atoms of neon (Ne), beryllium-like ($Ne^{6+}$) and helium-like ($Ne^{8+}$) atomic ions. At the same time, the main (leading in the infinite complete set) parts of the subthreshold resonance structure of the cross-section, the effects of radial relaxation of transition states in the $1s$-vacancy field, and only the projection $M=0$ of the total angular momentum $J=2$ of the wave function of the final state of ionization of *d*-symmetry were taken into account. In recent papers by the authors [12,13], the theory of papers [10,11] has been modified to take into account (a) the completeness of the set of virtual (intermediate) states of $1s \rightarrow np$ photoexcitation, (b) destructive quantum interference of partial amplitudes of the probability of radiation transition, and (c) the non-trivial complete ($J=2; M=0,\pm 1,\pm 2$) angular structure of the amplitude of the probability of radiation transition to the final states of ionization of *d*–symmetry. At the same time, a generalization of the results of the work [10,11] was given on the heavy neon-like ion of the iron atom ($Fe^{16+}$) [12] and the ions of the *isonuclear* sequence of the heavy nickel atom ($Ni^{18+} \rightarrow Ni^{24+} \rightarrow Ni^{26+}$) [13]. In this article, the theory of work [12,13] is modified (the dependence of the radiation decay widths of virtual $1snp$ photoexcitation states on the main quantum number is taken into account) and generalized to the *isoelectronic* sequence of heavy beryllium-like atomic ions. The objects of study are beryllium-like ions of titanium ($Ti^{18+}$; the charge of the nucleus of the ion $Z = 22$; the configuration and term of the ground state $[0] = 1s^2 2s^2 [{}^1S_0]$), chromium ($Cr^{20+}$; $Z = 24$), iron ($Fe^{22+}$; $Z = 26$) and zinc ($Zn^{26+}$; $Z = 30$). The choice of the isoelectronic sequence is due to the spherical symmetry of the ground state of the ions, their availability in the gas phase [14,15] during the experiment on the absorption of two XFEL-photons of energy $\hbar\omega$ ($\hbar$ – Planck's constant, $\omega$ – circular frequency of the photon) by an ion trapped in the "trap" [16,17] and the demand for their spectral characteristics, in particular, in X-ray diagnostics of hot laboratory [18,19] and astrophysical [20–22] plasma, as well as structural materials in controlled thermonuclear fusion facilities [23–25].

## 2. Theory

Consider the following channel of two-photon resonant single ionization of the K-shell of beryllium-like ion ions:

$$2\hbar\omega + [0] \rightarrow 1s(n,x)p + \hbar\omega \rightarrow 1s\varepsilon l. \qquad (1)$$

B (1) $n$ is the principal quantum number of excited states of the discrete spectrum, is the $x(\varepsilon)$ energy of the electron of the continuous spectrum, $l = s, d$ and the filled $2s^2$-shell is not specified. The structure of the presented ionization channel corresponds to the following approximations. First. The strong spatial and energetic separation of the valence $2s^2$-shell from the deep $1s^2$-shell of the ion makes it possible to neglect the production of the final $2s\varepsilon(s,d)$-states of two-photon ionization. For example, for the ion $Ti^{18+}$, the following inequalities are satisfied: $r_{1s} = 0.0367$ Å $<< r_{2s} = 0.1552$ Å ($r_{ns}$ – average radius $ns$ – shell of the ion backbone), $I_{1s} = 6019.25$ eV (calculation of this work) $>> I_{2s} = 1346.89$ eV [26] ($I_{ns}$ – energy of the ionization threshold $ns$ – shell of the ion backbone). Second. The probability amplitude of two-photon ionization along the channel $2\hbar\omega + [0] \rightarrow 1s\varepsilon l$ is determined by the contact interaction operator $\hat{C} \sim \sum_{n=1}^{N}(\hat{A}_n \cdot \hat{A}_n)$ ($N$ is the number of electrons in the ion) [27] and is proportional to the matrix element $\langle 1s | j_l | \varepsilon l \rangle$, where $j_l$ is the spherical Bessel function. For the electromagnetic field operator $\hat{A}_n$ (in the secondary quantization representation), the dipole approximation is assumed:

$$\hat{A}_n \rightarrow \sum_{\mathbf{k}} \sum_{\rho=1,2} \mathbf{e}_{\mathbf{k}\rho} (\hat{a}^+_{\mathbf{k}\rho} + \hat{a}^-_{\mathbf{k}\rho}). \qquad (2)$$

The structure of operator (2) corresponds to the fulfillment of the criterion of applicability of the dipole approximation:

$$(\mathbf{k} \cdot \mathbf{r}_n) << 1 \Rightarrow \exp[\pm i(\mathbf{k} \cdot \mathbf{r}_n)] \cong 1, \qquad (3)$$

$$r_{1s}/\lambda_\omega << 1. \qquad (4)$$

B (2–4) $\mathbf{e}_{\mathbf{k}\rho}$ ($\mathbf{k}$) is the polarization vector (wave vector) of the photon, $\hat{a}^+_{\mathbf{k}\rho}$ ($\hat{a}^-_{\mathbf{k}\rho}$) is the operator of the photon's production (annihilation), $\mathbf{r}_n$ is the radius-vector $n$ of the electron, and $\lambda_\omega$ is the wavelength of the absorbed photon. For example, for the ion $Ti^{18+}$ we have: $r_{1s} = 0.0367$ Å, $\lambda_\omega = 2.0330$ Å ($\hbar\omega = 6.10$ keV) and $r_{1s}/\lambda_\omega = 0.018 << 1$. Then $j_0 \rightarrow 1$, $j_2 \rightarrow 0$ and $\langle 1s | j_l | \varepsilon l \rangle \rightarrow 0$ due to the orthogonality of the radial parts of the wave functions $1s$– and $\varepsilon l$ –states. It should be noted here that in the recent work of the authors [28], the concept of the "criterion for the applicability of the dipole approximation" is modified. As a result, it is shown that the short-wave range of applicability of the dipole approximation (discarding, first of all, quadrupole corrections to the radiation transition operator) in terms of the energy of the absorbed photon is much wider than the region determined by the inequality (4).

## 2.1. Amplitude of ionization probability

The probability amplitude of two-photon ionization along channel (1) is physically interpreted in Fig. 1 in the formalism of Feynman diagrams [27] within the second (by the number of interaction vertices) order of the non-relativistic quantum perturbation theory. Let us establish its analytical structure. At the same time, the following formulas of this Section of the article are given in atomic units (a.u.; $e = \hbar = m_e = 1$). According to Fig. 1, for the desired amplitude, we have quantum interference of partial amplitudes of the probability of photoabsorption over the intermediate states of the discrete and continuous spectrum of *p*-symmetry:

$$A_l = A_l^{(1)} + A_l^{(2)}, \tag{5}$$

$$A_l^{(1)} = \sum_{M'} \sum_{N=2}^{\infty} \frac{\langle 0|\hat{R}|\Phi_n\rangle\langle\Phi_n|\hat{R}|\Psi_{el}\rangle}{\omega - I_{1snp} + i\gamma_{1s,n}}, \tag{6}$$

$$A_l^{(2)} = \sum_{M'} \int_0^\infty dx \frac{\langle 0|\hat{R}|\Phi_x\rangle\langle\Phi_x|\hat{R}|\Psi_{el}\rangle}{\omega - I_{1s} - x + i\gamma_{1s}}, \tag{7}$$

$$|0\rangle = [0] \otimes (a_\omega^+)^2 |0_{ph}\rangle, \tag{8}$$

$$|\Phi_{n,x}\rangle = |1s(n,x)p(^1P_1), M'\rangle \otimes \hat{a}_\omega^+ |0_{ph}\rangle, \tag{9}$$

$$|\Psi_{el}\rangle = |1s\varepsilon l(^1l_{J=l}), M\rangle \otimes |0_{ph}\rangle, \tag{10}$$

$$\hat{R} = -\frac{1}{c}\sum_{n=1}^{N}(\hat{p}_n \cdot \hat{A}_n). \tag{11}$$

In (6) – (11) the following are determined: the full wave functions of the initial ($|0\rangle$), intermediate ($|\Phi\rangle$) and final ($|\Psi\rangle$) states of two-photon ionization, $\hat{R}$ – the operator of the radiation transition, $c$ – the speed of light in a vacuum, $\hat{p}_n$ – the operator of the momentum of the *n*-electron, the projections of the total angular momentum of the system "ionic remnant ⊕ electron" $M' = 0, \pm 1$, $M = 0$ for $l = s$, $M = 0, \pm 1, \pm 2$ for $l = d$, $|0_{ph}\rangle$ is the wave function of the photon vacuum of quantum electrodynamics, $I_{1snp}$ is the photoexcitation energy, $\gamma_{1s,n} = \Gamma_{1s,n}/2$ and $\gamma_{1s} = \Gamma_{1s}/2$, where $\Gamma_{1s,n}$ ($\Gamma_{1s}$) is the width of decay $1s$ – vacancies of the virtual $1snp$ ($1sxp$)–state. Note that the $^1l_{J=l}$ structure of the final states of two-photon ionization ($J = 0 \Rightarrow {}^1S_0; J = 2 \Rightarrow {}^1D_2$) in (10) reproduces the Landau–Young theorem [29,30] for the total angular momentum of a system of two absorbed photons $J_\omega = 0, 2$. Using the methods of algebra of photon production (annihilation) operators [31], the theory of irreducible tensor operators [32], the theory of nonorthogonal orbitals [33] and assuming the approximation of plane waves, $|x(r)\rangle \sim \sin(r\sqrt{2x})$, for the single-electron amplitude of the probability of the radiation transition between continuum-spectrum states in (7),

$$(x-\varepsilon)\langle xp_+|\hat{r}|\varepsilon l_+\rangle \cong i\sqrt{2x} \cdot \delta(x-\varepsilon), \tag{12}$$

for the amplitude (5), we get:

$$A_s = \zeta\left(\mu + \sum_{n=2}^{\infty}\beta_n\langle np_+|\hat{r}|\bar{\varepsilon}s_+\rangle\right), \tag{13}$$

$$A_d = \sqrt{6}\zeta\left(\mu + \sum_{n=2}^{\infty}\beta_n\langle np_+|\hat{r}|\bar{\varepsilon}d_+\rangle\right) \cdot Q_M, \tag{14}$$

$$\mu = i \cdot 2\sqrt{2\bar{\varepsilon}}\,\langle 1s_0\|\hat{r}\|\bar{\varepsilon}p_+\rangle, \tag{15}$$

$$\beta_n = \frac{I_{1snp}(2\omega - I_{1snp})}{\omega - I_{1snp} + i\gamma_{1s,n}}\langle 1s_0\|\hat{r}\|np_+\rangle, \tag{16}$$

$$Q_M = -\frac{4\pi}{3}\sum_{M'}\sum_{p}(-1)^{M'}Y_{1,M'}(\mathbf{e}_\omega)Y_{1,p}^*(\mathbf{e}_\omega)\begin{pmatrix}1 & 1 & 2\\ -M' & p & M\end{pmatrix}. \tag{17}$$

Here $\delta$ is the Dirac delta-function, $\zeta = 4\pi/(3V\omega)$, $V(\text{cm}^3) = c$ is the volume of quantization of the electromagnetic field (numerically equal to the speed of light in a vacuum) [34], $\bar{\varepsilon} = 2\omega - I_{1s}$, $Y_{\alpha,\beta}(\mathbf{e}_\omega)$ is a spherical function, $p = 0, \pm 1$, "*" is the symbol of complex conjugation, the $3j$-Wigner symbol is determined, and the single-electron amplitude of the probability of photoexcitation $1s \to np$ is as follows:

$$\langle 1s_0 \| \hat{r} \| np_+ \rangle = \langle 1s_0 | 1s_+ \rangle \langle 2s_0 | 2s_+ \rangle^2 \langle 1s_0 | \hat{r} | np_+ \rangle, \tag{18}$$

$$\langle 1s_0 | \hat{r} | np_+ \rangle = \int_0^\infty P_{1s_0}(r)\, r\, P_{np_+}(r)\, dr. \tag{19}$$

In (12)–(16), (18), and (19), the indices "0" and "+" correspond to the $P(r)$ – radial part of the wave functions of electrons obtained by solving the single-configuration equations of the self-consistent Hartree–Fock field for the [0]–, $1s_+(n,x)p_+$ – and $1s_+\varepsilon l_+$ – ion configurations.

## 2.2. Generalized ionization cross-section

Let us establish the analytical structure of the generalized cross-section of a two-photon resonant single ionization of an atomic ion. Following the definition of the concept of such a cross-section [35],

$$d\sigma_g^{(l)} = (V/2c)\, d\sigma_l, \tag{20}$$

considering Fermi's "golden rule" [36],

$$d\sigma_l = (\pi V/c)|A_l|^2 \cdot \delta(\varepsilon - \bar{\varepsilon})\, d\varepsilon, \tag{21}$$

integrating into (21) by the energy of the photoelectron and summing up by $l = s, d$ (by the probabilities of physically different final states of ionization), for the desired total generalized cross-section, we obtain (the probability of photon disappearance without photoelectron registration):

$$\sigma_g(\text{cm}^4 \cdot \text{s}) = (\eta/\omega)^2 \sum_{l=s,d}\sum_{i=1,2} a_l \cdot L_{il}^2, \tag{22}$$

$$L_{1l} = \sum_{n=2}^\infty (\omega - I_{1snp})\, B_{ln}, \tag{23}$$

$$L_{2l} = D - \sum_{n=2}^\infty \gamma_{1s,np}\, B_{ln}, \tag{24}$$

$$D = 2\sqrt{2\bar{\varepsilon}} \cdot \langle 1s_0 \| \hat{r} \| np_+ \rangle, \tag{25}$$

$$B_{ln} = \frac{\beta_n}{\omega - I_{1snp} - i\gamma_{1s,n}} \langle np_+ | \hat{r} | \bar{\varepsilon} l_+ \rangle. \tag{26}$$

Here it is taken into account that for virtual $xp$-states of a continuous spectrum, the energy denominator of the probability amplitude of $1s \to xp$ photoionization at the point of the energy pole $\bar{\varepsilon} = 2\omega - I_{1s}$ takes the form $\omega - i\gamma_{1s} = \omega \cdot (1 - i\gamma_{1s}/\omega) \cong \omega$ at $\omega \sim I_{1s} \gg \gamma_{1s}$. Here, too, the values $\eta(\text{cm}^4 \cdot \text{s}) = 0.278 \cdot 10^{-52}$, $a_s = 1$, $a_d = \frac{3}{2}a_d^{(0)}\left(1 - \frac{1}{4\pi}\right)$ and $a_d^{(0)} = 4/5$ [10,11] (only the projection $M$

= 0 total angular momentum $J = 2$ is taken into account) are defined. The value of the coefficient $a_d$ is obtained according to the formula

$$a_d = 6 \cdot \sum_{M=-2}^{2} |Q_M|^2 \tag{27}$$

taking into account the analytical result of the work [37] for the sum of the products of two 3$j$-Wigner symbols. At the same time, the scheme of the proposed XFEL-experiment for linearly polarized absorbed photons is implemented: $\mathbf{k} \in OZ$, $\mathbf{e}_\omega \in OX$ ($OX$, $OZ$ – axes of the rectangular coordinate system),

$$Y_{1,0}(\mathbf{e}_\omega) = 0, \quad Y_{1,\pm 1}(\mathbf{e}_\omega) = \mp 3/(4\pi\sqrt{2}). \tag{28}$$

Note here that $(a_d^{(0)}/a_d) \cdot 100\,(\%) \cong 72\,(\%)$. Thus, the additional consideration of projections $M = \pm 1, \pm 2$ of the total angular momentum $J = 2$ significantly (by ~ 30%) increases the contribution of the generalized cross-section component (22) for $l = d$.

In (6), the following approximation of the total decay widths $1s$-vacancies of photoexcitation states $1s \to np$ is adopted:

$$\Gamma_{1s,n} = \Gamma_A + \Gamma_{R,n}, \tag{29}$$

$$\Gamma_{R,n} = \alpha n^{-(\beta + \gamma/n)}, \tag{30}$$

$$\lim_{n \to \infty} \Gamma_{1s,n} = \Gamma_A. \tag{31}$$

In (29) $\Gamma_A$ is the Auger width of the channel decay $1s2s^2 \to 1s^2 \varepsilon s$ in the approximation of independence $\Gamma_A$ from the principal quantum number ($np$ the electron plays the role of an "observer") and $\Gamma_{R,n}$ is the width of the radiative decay along the channel $1snp \to 1s^2$. The parameters $\Gamma_A$, $\alpha$, $\beta$, are $\gamma$ determined by interpolation (for $Ti^{18+}$ and $Cr^{20+}$) and extrapolation (for $Zn^{26+}$) of the theoretical data of the work [38] for $n = 2$ in $Mg^{8+}$, $Ar^{14+}$, $Fe^{22+}$ ions, taking into account the recurrence formula [39]:

$$\frac{\Gamma_{R,n}}{\Gamma_{R,n+1}} = \left(\frac{I_{1snp}}{I_{1s(n+1)p}}\right)^3 \left(\frac{\langle 1s|\hat{r}|np \rangle}{\langle 1s|\hat{r}|(n+1)p \rangle}\right)^2, \quad n \geq 2. \tag{32}$$

With large values $n$ we have:

$$\Gamma_{R,n} \sim n^{-\beta}. \tag{33}$$

Formula (33) qualitatively reproduces (see $\beta \cong 3$ in Section 3 of this article) is the result of the quantum defect theory $\Gamma \sim n^{-3}$ [40] with the difference that in [40] the widths of the autoionization resonances of the photoabsorption cross-sections are investigated.

Formally, mathematically infinite functional series in (23), (24) correspond to the consideration of the completeness of the set of virtual states of photoexcitation $1s \to np$. For the hydrogen atom, hydrogen-like atomic ions, and multielectron atoms, methods of *analytic* summation of such series are effective in approximating the *zero* width of the decays of the spanning vacancies (the energy denominator in the Coulomb Green function contains the term $i\gamma \to i0$ [41]) – methods of the Coulomb Green's function [42], "average frequency" in the energy denominators of probability amplitudes of multiphoton ionization [43], inhomogeneous differential equations for perturbed (correlation) functions [44–46], and "stable variation" for probability amplitudes of multiphoton processes [47–49]. However, as far as we know, the problem of analytical accounting of the completeness of the set for an atom (atomic ion) with *non-zero* decay widths, first of all, inner shell vacancies, remains open. In this paper, the numerical method of summation of authors is implemented [50]. The values $I_{1snp}$ and $J_n = \langle 1s_0 \| r \| np_+ \rangle$ for $n \in [2; 10]$ are obtained in the Hartree-Fock single-configuration approximation. For $n \in [11; \infty)$ photoexcitation energies $1s \to np$ are obtained by approximation of the form:

$$I_{1snp} = I_{1s} - \frac{1}{n^2}\left(a - \frac{b}{n}\right), \quad \lim_{n\to\infty} I_{1snp} = I_{1s}, \tag{34}$$

where the parameters $a$ and $b$ are defined by the values $I_{1smp}$ for $m = 9, 10$. For $n \in [11; \infty)$ the amplitudes of the probability of photoexcitation $1s \to np$ are obtained by approximation of the form:

$$J_n = \frac{1}{n^2}\left(c + \frac{d}{n} + \frac{f}{n^2}\right), \quad \lim_{n\to\infty} J_n = 0, \tag{35}$$

where the parameters $c$, $d$ and $f$ are defined by the values $J_m$ for $m = 8, 9, 10$. For the integral $\langle np_+|\hat{r}|\bar{\varepsilon}l_+\rangle$ in (26), the formula (35) is realized, taking into account that the parameters $c$, $d$ and $f$ become functions of the energy of the absorbed photon (see $\bar{\varepsilon} = 2\omega - I_{1s}$). At the same time, it should be taken into account that, as a rule, numerical solutions of the equations of a self-consistent Hartree-Fock field do not satisfy the physical requirement of orthogonality of wave functions of $|1s_+\rangle$ – and $|2s_+\rangle$ –states to an excited $|\bar{\varepsilon}s_+\rangle$ –state of a continuous spectrum *of the same* symmetry:

$$\alpha = \langle 1s_+|\bar{\varepsilon}s_+\rangle \neq 0, \quad \beta = \langle 2s_+|\bar{\varepsilon}s_+\rangle \neq 0. \tag{36}$$

To restore orthogonality, this article adopts the Gram-Schmidt orthogonalization procedure [51]:

$$|\bar{\varepsilon}s_+\rangle \to |\tilde{\varepsilon}s_+\rangle = |\bar{\varepsilon}s_+\rangle - \alpha \cdot |1s_+\rangle - \beta \cdot |2s_+\rangle, \tag{37}$$

$$\langle 1s_+|\tilde{\varepsilon}s_+\rangle = \langle 2s_+|\tilde{\varepsilon}s_+\rangle = 0. \tag{38}$$

As a result, the desired integral in (26) takes the form:

$$\langle np_+|\hat{r}|\tilde{\varepsilon}s_+\rangle = \langle np_+|\hat{r}|\bar{\varepsilon}s_+\rangle - \alpha\langle np_+|\hat{r}|1s_+\rangle - \beta\langle np_+|\hat{r}|2s_+\rangle. \tag{39}$$

The results of Fig. 2 demonstrate a significant role of the "orthogonalization effect" in constructing the amplitudes of the probability of a radiative transition $np_+ \to \tilde{\varepsilon}s_+$. A mathematically more rigorous approach (the concept of the extended Hilbert space) to the construction of probability amplitudes of radiation transitions with the implementation of the Gram-Schmidt orthogonalization procedure is presented in the authors' paper [52].

## 3. Results and discussion

The calculation results are shown in Fig. 3 – 7 and in Table 1 – 3. The values of the parameters of the generalized cross-section (22) used in the calculations are given in Table 1 and Fig. 3. For the energies of XFEL-photons the range is $\hbar\omega \in (3; 13)$ keV ([53] XFELO, USA; [54,55] European XFEL, Germany; [56] PAL–XFEL, Republic of Korea).

The results in Figs. 4 – 7 and Table 2 demonstrate a pronounced subthreshold resonance structure of the generalized cross-sections of the two-photon single ionization of the ions under study, due to virtual states of photoexcitation $1s \to np$ (the values of the main quantum number $n \in [2; 500]$ are taken into account). The decrease in the absolute values of the generalized cross-section resonances for a number of ions under study is due to a decrease $\langle 1s_0|\hat{r}|np_+\rangle-$, $\langle np_+|\hat{r}|\tilde{\varepsilon}(s,d)_+\rangle-$ amplitudes of the probability of radiation transitions in (26) and an increase in the total $\Gamma_{1s,n}-$widths of decay $1snp$ –states of photoexcitation (see Fig. 3). For example, for the leading photoexcitation resonance we have: $\langle 1s_0|\hat{r}|2p_+\rangle \cdot 10^2 = 5.8$ (Ti$^{18+}$), 4.3 (Zn$^{26+}$), $\langle 2p_+|\hat{r}|\varepsilon d_+\rangle \cdot 10^3 \left(\frac{1}{\sqrt{a.u.}}\right) = 1.5$ (Ti$^{18+}$), 0.8 (Zn$^{26+}$) and $\Gamma_{1s,2}(eV) = 0.225$ (Ti$^{18+}$), 0.879 (Zn$^{26+}$).

The results in Figs. 4 – 7 also demonstrate the effect of destructive (quenching) quantum interference of the amplitudes of the probability of radiation transitions $1s \to np$. This effect is due to the alternating nature of the multiplier $\hbar\omega - I_{1snp}$ in (23) and the "immersion" of these states in the continuum (see $D$ from (24)). At the same time, for a number of ions under study, (a) energy distances increase between the maxima of resonances of the generalized cross-section and (b) "windows of transparency" appear in the form of a sharp drop in the probability of two-photon ionization.

For the single-electron amplitudes of the bremsstrahlung absorption probability in (26), the inequality is met:

$$\langle np_+ | \hat{r} | \bar{\varepsilon} d_+ \rangle > \langle np_+ | \hat{r} | \bar{\varepsilon} s_+ \rangle . \tag{40}$$

For example, for the leading photoexcitation resonance in the Ti$^{18+}$ ion, we have $\langle 2p_+ | \hat{r} | \bar{\varepsilon} d_+ \rangle$ $10^3 (\frac{1}{\sqrt{a.u.}}) = 1.5 > \langle 2p_+ | \hat{r} | \bar{\varepsilon} s_+ \rangle 10^3 (\frac{1}{\sqrt{a.u.}}) = 0.5$. As a result, the data in Table 3 show the leading role of the $d$-symmetry of the final ionization state in determining the absolute values, first of all, the resonances of the total generalized cross-section at $\hbar\omega \in (3; 13)$ keV.

## 4. Conclusion

A non-relativistic version of the quantum theory of the process of two-photon resonant single ionization of the K-shell of a heavy beryllium-like atomic ion has been constructed. The isoelectron sequence of ions has been studied: Ti$^{18+} \to$ Cr$^{20+} \to$ Fe$^{22+} \to$ Zn$^{26+}$. A pronounced resonant subthreshold structure of the generalized cross-section and the effect of destructive quantum interference of the probability amplitudes of radiation transitions into virtual states of $p$-symmetry were established. The leading role of the $d$-symmetry of the final ionization state in determining the absolute values of the total generalized cross-section in the energy region of XFEL-photons $\hbar\omega \in (3; 13)$ keV was also established. Taking into account correlation and relativistic effects and going beyond the dipole approximation for $\hat{R}(\hat{C})$ – operator of the radiation (contact) transition is the subject of future development of the theory. The results of successful experiments on the observation of two-photon ionization of atoms, molecules and solids (see, e.g., [59,60] and references therein) suggest that the absolute values of the generalized cross-sections in Fig. 4 – 7 are quite measurable in the modern XFEL-experiment.

**Conflict of interest**

The authors state that they have no conflict of interest.

**Table 1.** Parameters of total decay widths $1snp$ – photoexcitation states (obtained from the theoretical data of the work [38]) from formula (29) and the energy of ionization thresholds (calculation of this work) $1s^2$– shells ($I_{1s}$) of the ions under study.

| Ion | $\Gamma_A$, eV | $\alpha$, eV | $\beta$ | $\gamma$ | $I_{1s}$, keV |
|---|---|---|---|---|---|
| Ti$^{18+}$ | 0.085 | 2.568 | 3.282 | 1.831 | 6.019 |
| Cr$^{20+}$ | 0.087 | 3.474 | 3.279 | 1.679 | 7.230 |
| Fe$^{22+}$ | 0.091 | 5.062 | 3.271 | 1.686 | 8.553 |
| Zn$^{26+}$ | 0.097 | 13.051 | 3.264 | 1.589 | 11.544 |

**Table 2.** Spectral characteristics of the leading photoexcitation resonances $1s \rightarrow np$ of generalized cross-sections of a two-photon resonant single ionization of the K-shell of the ions under study in the energy region of the absorbed photons $\hbar\omega \in (3; 13)$ keV.

| Ion | $np$ | $I_{1snp}$, keV | $\sigma_g$, $10^{-52}$ cm$^4$·s |
|---|---|---|---|
| Ti$^{18+}$ | 2p | 4.6752 | 4.5445 |
|  | 3p | 5.4481 | 0.5120 |
|  | 4p | 5.7020 | 0.0979 |
| Cr$^{20+}$ | 2p | 5.5998<br>5.6160$^a$ | 2.3279 |
|  | 3p | 6.5341 | 0.3405 |
|  | 4p | 6.8430 | 0.0723 |
| Fe$^{22+}$ | 2p | 6.6099<br>6.6287$^b$ | 1.1402 |
|  | 3p | 7.7212 | 0.1967 |
|  | 4p | 8.0904 | 0.0479 |
| Zn$^{26+}$ | 2p | 8.8897 | 0.2868 |
|  | 3p | 10.4015 | 0.0798 |
|  | 4p | 10.9078 | 0.0294 |

$^a$ Experimental of work [57].
$^b$ Relativistic calculation of work [58].

**Table 3.** The relative contribution of $s-$ and $d-$ symmetries of the final ionization state $\Lambda = \sigma_g^{(d)}/\sigma_g^{(s)}$ (see (22) for the terms $l=d$, $l=s$) to the total generalized cross-section of the two-photon resonant single ionization of the K-shell of the ions under study. The energies of the absorbed photons $\hbar\omega = I_{1s2p}$ (*top line*), $I_{1s}+1$(keV) (*bottom line*) are taken.

| Ion | $\hbar\omega$, keV | $\Lambda$ |
|---|---|---|
| Ti$^{18+}$ | 4.6752 | 8.4972 |
|  | 7.0200 | 1.3041 |
| Cr$^{20+}$ | 5.5998 | 10.3713 |
|  | 8.2300 | 1.3753 |
| Fe$^{22+}$ | 6.6099 | 8.5061 |
|  | 9.5530 | 1.4242 |
| Zn$^{26+}$ | 8.8897 | 9.9992 |
|  | 12.5440 | 1.5705 |

**Fig. 1.** Probability amplitude of two-photon resonant single ionization of the K-shell of a beryllium-like atomic ion in the representation of Feynman diagrams ($l=s,d$). Time direction – left to right ($t_1 < t_2$). Right arrow – electron, left arrow – vacancy. Double line – state obtained in the Hartree-Fock field of $1s-$ vacancy. $\hbar\omega$ is an absorbed photon.

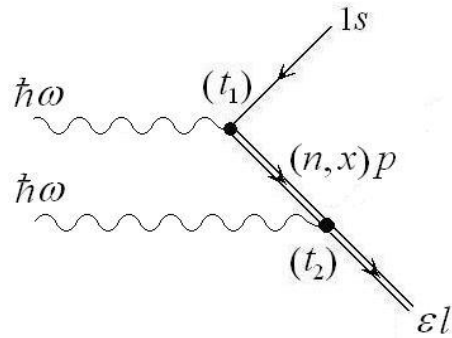

**Fig. 2.** "Orthogonalization effect" in constructing the amplitude of the probability of radiation transition (bremsstrahlung absorption) $np_+ \to \tilde{\varepsilon}s_+$ in the $Cr^{20+}$ ion: **1** – before orthogonalization; **2** – after orthogonalization. $\tilde{\varepsilon} = 2I_{1snp} - I_{1s}$.

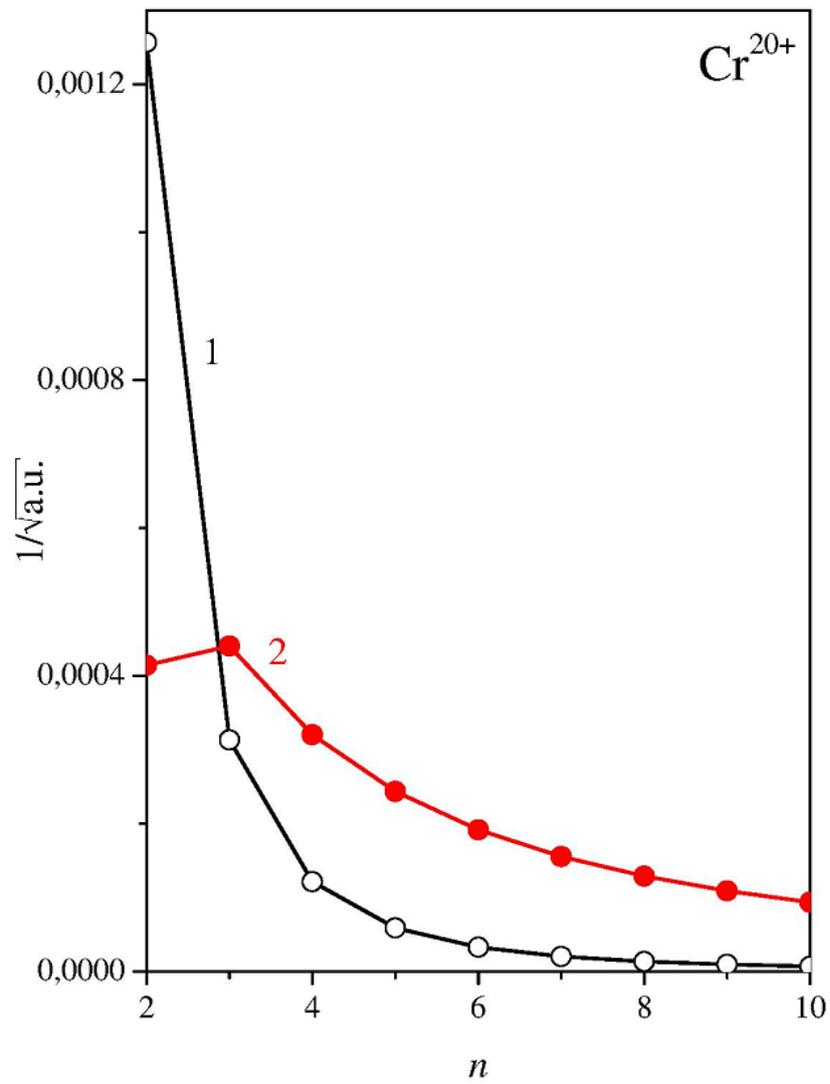

**Fig. 3.** Total decay widths of photoexcitation states $1s \to np$ ($\Gamma_{1s,n}$) for ions $Zn^{26+}$ (**1**), $Fe^{22+}$ (**2**), $Cr^{20+}$ (**3**) and $Ti^{18+}$ (**4**), calculated according to the formula (29).

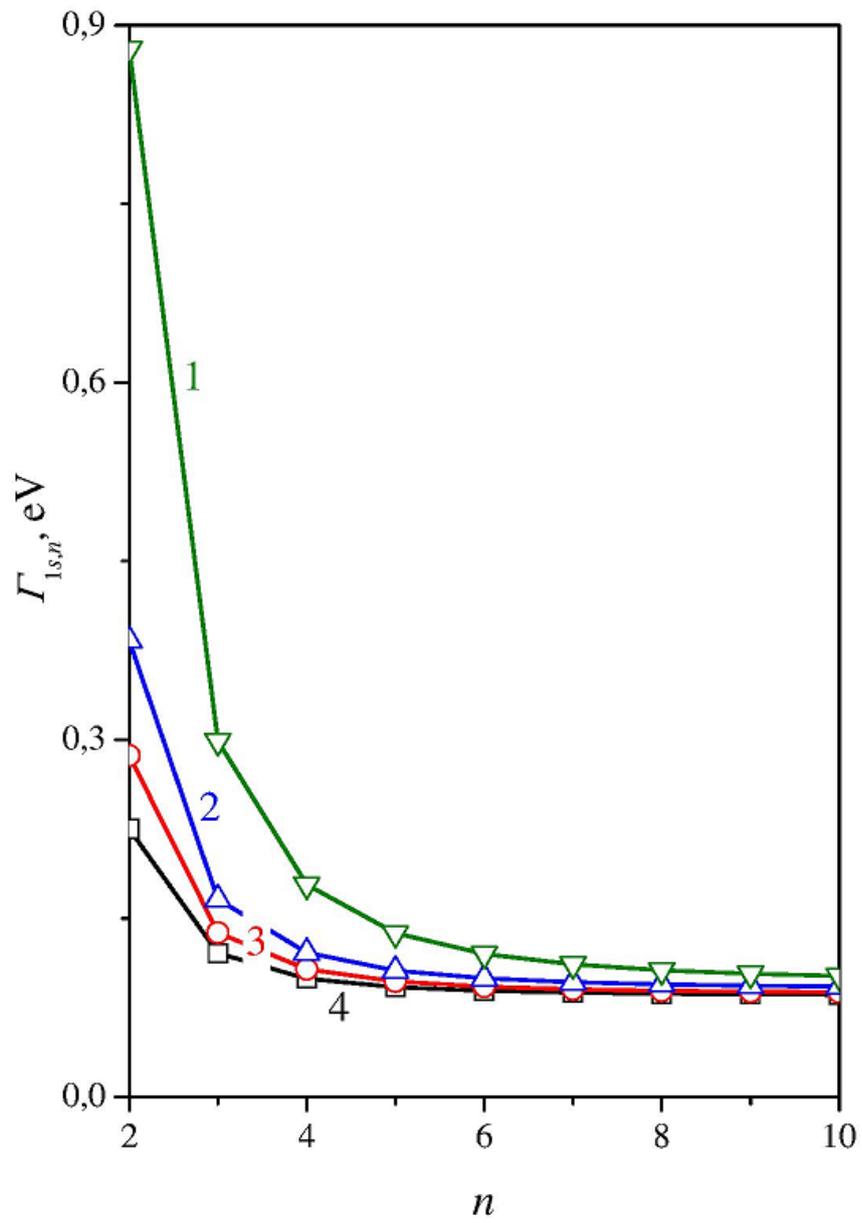

**Fig. 4.** Total generalized cross-section of two-photon resonant single ionization of K-shell of ion Ti$^{18+}$. $\hbar\omega$ is an absorbed photon.

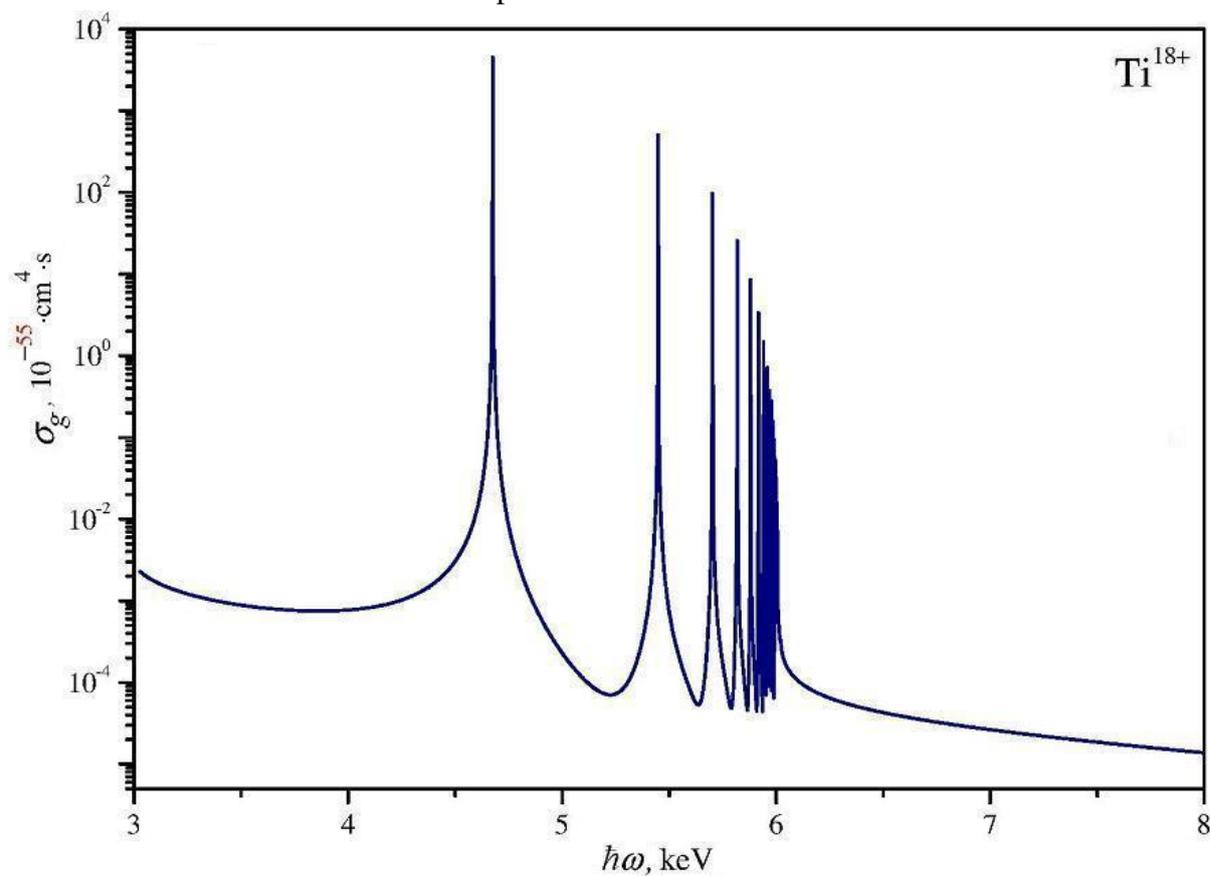

**Fig. 5.** See Fig. 4, but for ion Cr$^{20+}$.

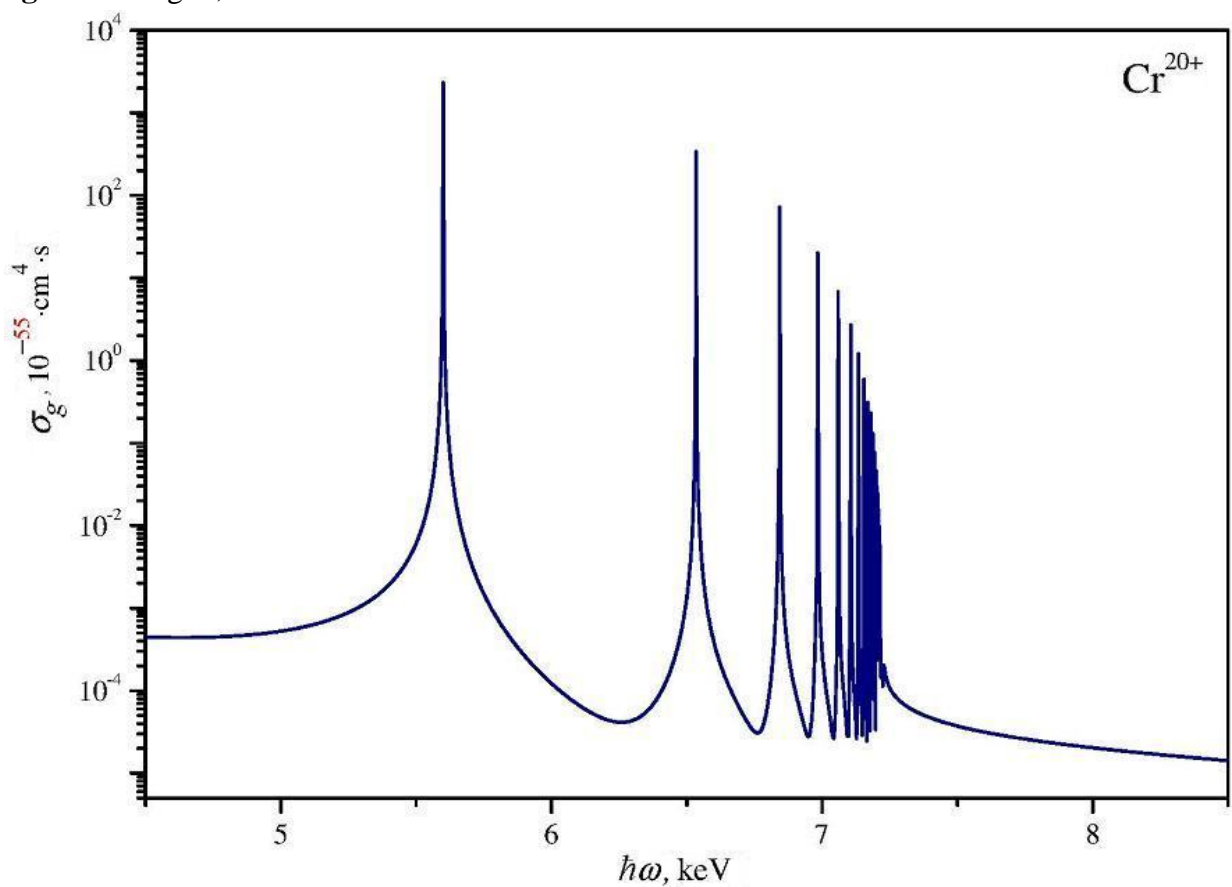

**Fig. 6.** See Fig. 4, but for ion $Fe^{22+}$.

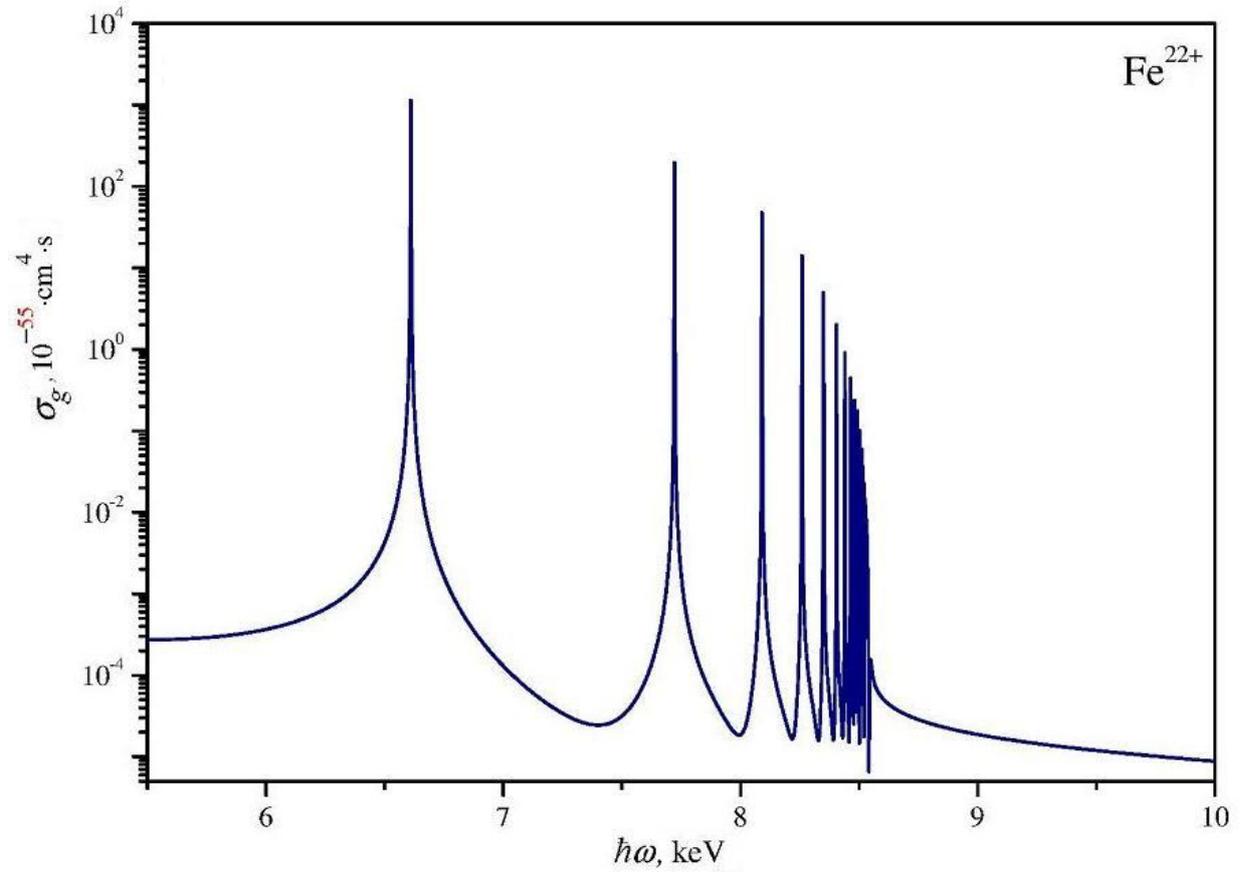

**Fig. 7.** See Fig. 4, but for ion $Zn^{26+}$.

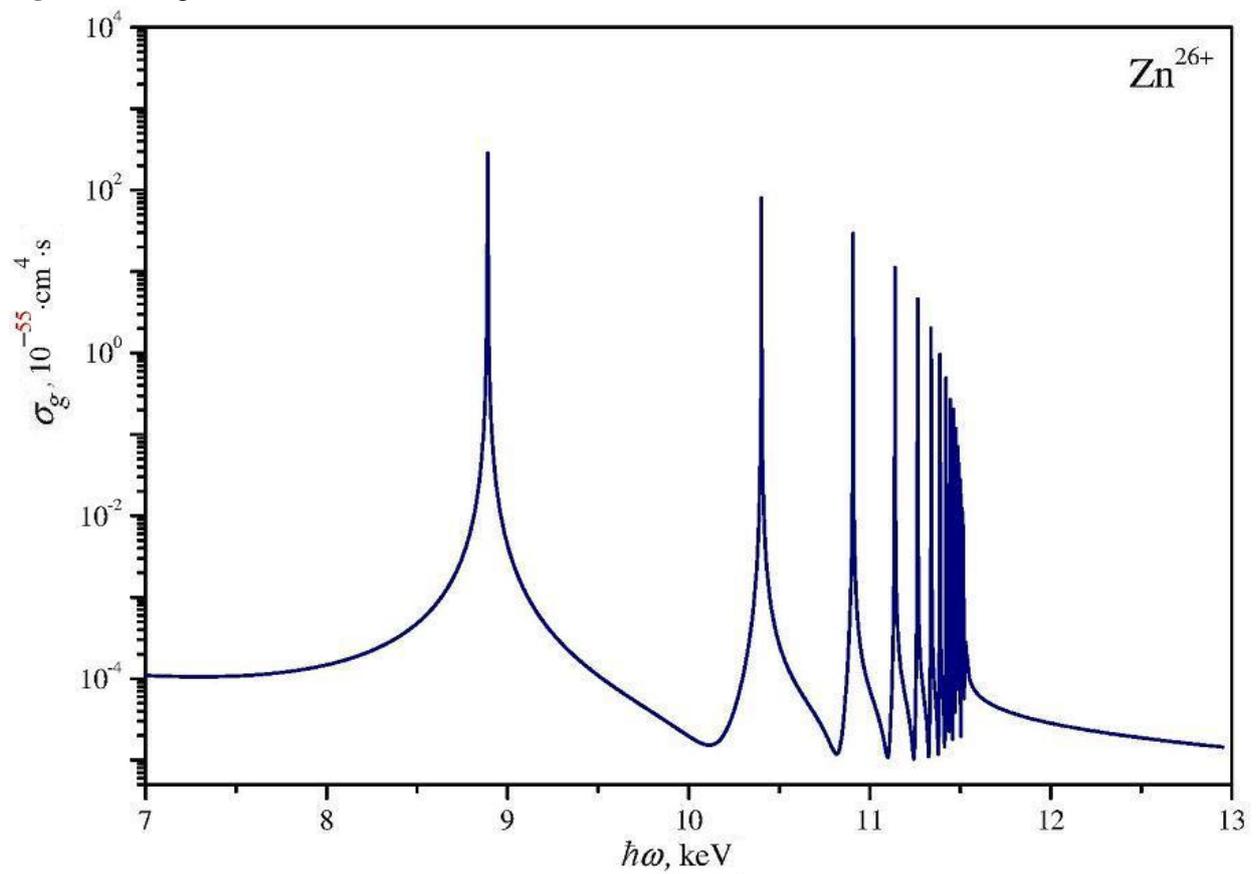